\documentclass[11pt]{article} 
\usepackage{graphicx}
\newcommand{\be}{\begin{equation}}
\newcommand{\ee}{\end{equation}}
\newcommand{\bea}{\begin{eqnarray}}
\newcommand{\eea}{\end{eqnarray}}
\newcommand{\sn}{{\rm sn}}

\newcommand{\dn}{{\rm dn}}
\newcommand{\cn}{{\rm cn}}
\newcommand{\sech}{{\rm sech}}

\begin{document}

\vspace{0.5in}
\begin{center}
{\LARGE{\bf One-Parameter Family of Elliptic Sine-Gordon Equations}}
\end{center}

\begin{center}
{\LARGE{\bf Avinash Khare}} \\
{Physics Department, Savitribai Phule Pune University \\
 Pune 411007, India}
\end{center}

\begin{center}
{\LARGE{\bf Avadh Saxena}} \\ 
{Theoretical Division and Center for Nonlinear Studies, 
Los Alamos National Laboratory, Los Alamos, New Mexico 87545, USA}
\end{center}

\vspace{0.9in}
\noindent{\bf {Abstract:}}

We introduce a continuous one-parameter family of elliptic sine-Gordon 
equations (SGE) characterized by the modulus $0 \le m \le 1$ of Jacobi
elliptic functions and analyze some of its properties and obtain its kink
solution for various values of modulus $m$. These elliptic SGE have the 
novel property that while in the limit $m = 0$ they go over to the 
integrable sine-Gordon equation, in the $m = 1$ limit they go over to the 
integrable sine hyperbolic-Gordon equations (SHGE). 

\section{Introduction}

Sine-Gordon (SG) equation is one of the most celebrated integrable equations 
which has been extensively studied for more than sixty years \cite{raj} and 
has found a large number of applications in several different areas of physics 
ranging from nonlinear pendulums to Josephson junctions \cite{cou}. Another 
well known integrable equation is the sine hyperbolic-Gordon (SHG) equation 
which is related to constant mean curvature surfaces \cite{shg}. It is then natural 
to enquire if one can find a one-parameter family of elliptic SG-SHG equations 
which will go over to the SG equation in one limit and the SHG equation in the 
other limit. This is the task that we have undertaken in this paper. We emphasize 
that what we call elliptic SG equation here is {\it different} from the commonly 
used term in the literature, which refers to a semilinear elliptic PDE typically 
with a special double well potential 

The purpose of this paper is to introduce a continuous one-parameter family of 
so called ``elliptic'' SG-SHG potentials of the form
\be\label{1}
V(\phi) = \frac{\cn(\phi,m)}{\dn^2(\phi,m)}\,,
\ee
where $\cn(x,m),\dn(x,m)$ and also $\sn(x,m)$ are three basic Jacobi elliptic
functions which depend on the continuous modulus parameter $0 \le m \le 1$ 
\cite{as}. We analyze some of the properties of the elliptic SG 
equations and obtain kink solutions for the entire range of
$0 < m < 1$, and show that the kink solutions have exponential tail at any value
of $0 < m < 1$ except at $m = 1/2$ where one has a kink solution with a power 
law tail. It is of course well known that the SG kink solution has an 
exponential tail while the SHG equation does not admit a kink solution (since 
it has a single well potential).  In addition, we point out a novel connection between 
the static solutions of the SG and the SHG equations. 

The plan of the paper is the following. In Sec. II we establish a novel connection 
between the static solutions of the SG and SHG equations. In Sec. III
we introduce and discuss in some detail one-parameter family of elliptic 
SG-SHG equations and obtain the kink and the antikink solutions for the 
entire class of one-parameter family of potentials.  Finally, in Sec. IV we 
summarize the main results obtained in this paper and point out some of the 
open problems. 

\section{Novel Connection Between the Static Solutions of SG and SHG Equations}

Let us start from the static SG equation
\be\label{2}
\phi_{xx} = \sin(\phi)\,.
\ee
On using the ansatz
\be\label{3}
\cos(\phi/2) = u\,,
\ee
the SG Eq. (\ref{2}) goes over to the nonlinear equation
\be\label{4}
(1-u^2)u_{xx}+u(u_x)^2 = -u(1-u^2)^2\,.
\ee
Let us now consider the static SHG equation
\be\label{5}
\phi_{xx} = \sinh(\phi)\,.
\ee
On using the ansatz
\be\label{6}
\cosh(\phi/2) = u\,,
\ee
we find that the SHG Eq. (\ref{5}) goes over to the same nonlinear 
Eq. (\ref{4}). Thus once we find solutions of the nonlinear Eq. (\ref{4}) 
then we simultaneously obtain the solutions of both the static SG Eq. (\ref{2})
and the static SHG Eq. (\ref{5}). We might note that normally the way one 
relates the static solutions of the two equations is by noting that as 
$\phi \rightarrow i\phi, \sin(\phi)$ goes over to $i\sinh(\phi)$ and the SG Eq.
(\ref{2}) goes over to the SHG Eq. (\ref{5}).

\section{Elliptic SG Potential and its Kink and Antikink Solutions}

Let us consider the equation
\be\label{1.1}
\phi_{tt} - \phi_{xx} + \frac{dV}{d\phi} = 0\,,
\ee
where
\be\label{1.2}
V(\phi) = \frac{\cn(\phi,m)}{\dn^2(\phi,m)}\,.
\ee
Note that in the limit $m = 0$, since $\dn(\phi,m=0) = 1, \cn(\phi,m=0) =
\cos(\phi)$ \cite{as}, hence
\be\label{1.3}
V(\phi,m= 0) = \cos(\phi)\,.
\ee
Thus at $m = 0$ we get the SG equation with a wrong sign i.e.
\be\label{1.4}
\phi_{tt} - \phi_{xx} -\sin(\phi) = 0\,. 
\ee
However, on using $\phi = \theta +\pi$, then in terms of $\theta$ we do get
the correct SG equation 
\be\label{1.5}
\theta_{tt} - \theta_{xx} +\sin(\theta) = 0\,.
\ee
On the other hand, in the limit $m = 1$, since $\dn(\phi,m=1) = \cn(\phi,m=1) 
= \sech(\phi)$ \cite{as}, hence
\be\label{1.5a}
V(\phi,m= 1) = \cosh(\phi)\,,
\ee
so that in the hyperbolic limit of $m = 1$, we do get the SHG equation
\be\label{1.6}
\phi_{tt} - \phi_{xx} +\sinh(\phi) = 0\,.
\ee

Before we try to obtain the kink solution for this general model, let us evaluate the 
maxima and the minima of the potential. It is easy to see that for the elliptic SG model
\bea\label{1.7}
&&V(\phi) = \frac{\cn(\phi,m)}{\dn^2(\phi,m)}\,, \nonumber \\
&&V'(\phi) = \frac{\sn(\beta \phi,m)}{\dn^4(\phi,m)}[\dn^2(\phi,m)
-2(1-m)]\,, \nonumber \\
&&V''(\phi) = \frac{(5-4m)cn(\phi,m)}{\dn^2(\phi,m)} 
-\frac{6(1-m)cn(\phi,m)}{\dn^4(\phi,m)}\,.
\eea
Thus the extrema of the potential are when 
\be\label{1.8}
\sn(\phi,m) = 0\,,~~\dn(\phi,m) = \sqrt{2(1-m)}\,.
\ee
Note $\sqrt{1-m} \le \dn(x,m) \le 1$ which implies that when $m < 1/2$ then the
only extrema are when $\sn(\phi,m) = 0$ while for $m > 1/2$ there are 
three extrema as given above. Further, when $m = 1/2$, $\dn(\phi,m) = 1$
so that the extrema are again when $\sn(\phi,m) = 0$. Thus we need to
consider the cases of $m \le 1/2$ and $m > 1/2$ separately.

{\bf Case I: $0 < m \le 1/2$}

In this case the two minima are at $\phi = \pm 2K(m)$ while the maximum is at 
$\phi = 0$ with $V_{mim} = -1$ while $V_{max} = +1$. Here $K(m)$ is the complete 
elliptic integral of the first kind \cite{as, gr}. Thus in this case there
should be a kink solution from $-2K(m)$ to +$2K(m)$ as $x$ goes from $-\infty$
to $\infty$.

A plot of the potential $V(\phi)$ as given by Eq. (\ref{1.7}) vs $\phi$ for $m = 1/4$, 
$m=1/2$ and $m=3/4$ is given in Figure 1. 

\begin{figure}[h] 
\includegraphics[width=5.0 in]{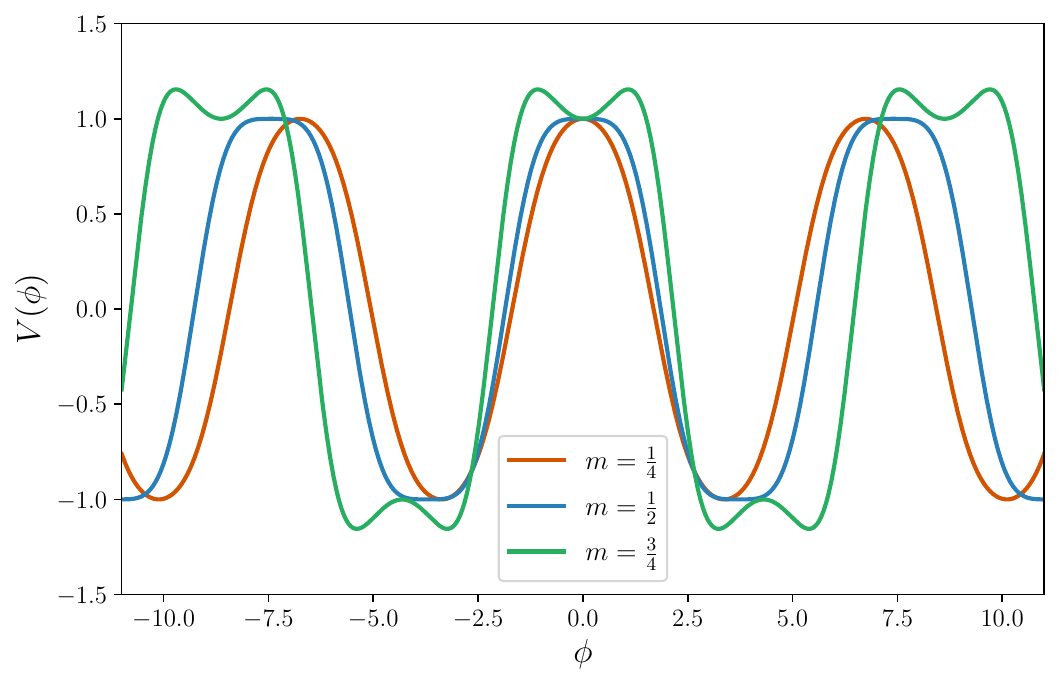}
\caption{Potential in Eq. (1) for three values of the elliptic function modulus $m$. 
Note that for $m >1/2$ the minima and maxima are split into two each (green curve) }
\end{figure}  

{\bf Case II: $m > 1/2$}

In this case the extrema are as given by Eq. (\ref{1.8}). On calculating $V''(\phi)$ 
at these extrema it turns out that in this case while the absolute minimum is
at $\cn(\beta,m) = -\sqrt{\frac{(1-m)}{m}}$, the local minimum is at $\phi = 0$
with $V(\phi = 0,m) = 1$ while $V[\cn(\phi,m) 
= - \sqrt{\frac{(1-m)}{m}}] = -\frac{1}{2\sqrt{m(1-m)}}$. On the other hand,
the maxima are at  $\phi = \pm 2K(m)$ and $\cn(\phi,m) 
= + \sqrt{\frac{(1-m)}{m}}$ with $V(\phi = \pm 2K(m),m) = -1$ while 
$V[\cn(\phi,m) = + \sqrt{\frac{(1-m)}{m}}] = +\frac{1}{2\sqrt{m(1-m)}}$.
Thus in this case there should be a kink solution from $-\phi_c$ to +$\phi_c$ 
as $x$ goes from $-\infty$ to +$\infty$ where $\pm \phi_c$ are such that 
$\cn(\pm \phi_c,m) = -\sqrt{\frac{(1-m)}{m}}$.


Note that the above analysis is not valid for 
$m = 0, 1$ but then we already know the kink solution at $m = 0$, and also know 
that there is no kink solution at $m = 1$. 

\subsection{Kink Solution For $0 < m < 1/2$}

Let us now try to obtain the kink solution for $0 < m < 1/2$. In order to 
get the kink solution, we start from the static self-dual equation
\be\label{1.9}
\phi_{x} = \sqrt{2V(\phi)} = \pm \sqrt{2\left[1+\frac{\cn(\phi,m)}{\dn^2(\phi,m)}\right]}\,,
\ee
where we have added a constant so that the potential is nonnegative. Thus
we need to solve the integral
\be\label{1.10}
\int \frac{d\phi}{\sqrt{[1+\frac{\cn(\phi,m)}{\dn^2(\phi,m)}]}} = -\sqrt{2} x\,.
\ee
On using $\dn^2(x,m) = 1-m + m\cn^2(x,m)$, the integral is given by
\be\label{1.11}
\int \frac{\dn(\phi,m) d\phi}{\sqrt{[1-m +m\cn^2(\phi,m)+\cn(\phi,m)]}} 
= -\sqrt{2} x\,.
\ee
On making the substitution $\cn(\phi,m) = y$, the integral becomes
\be\label{1.12}
\int \frac{dy}{\sqrt{(1-y^2)(1-m +m y^2+y)}} = \sqrt{2} x\,. 
\ee
Additionally, noting that $1-m+my^2+y = (1+y)[1-m(1-y)]$, the integral (\ref{1.12}) can
be reexpressed as
\be\label{1.13}
\int \frac{dy}{(1+y)\sqrt{(1-y)[1-m(1-y)]}} = \sqrt{2} x\,. 
\ee
On substituting $y+1 = u$ we have the integral of the form
\be\label{1.14}
\int \frac{du}{u\sqrt{a+bu+cu^2}} = \sqrt{2}x\,,
\ee
where in this case
\be\label{1.15}
a = 2(1-2m) > 0\,,~~b = 4m-1\,,~~c = -m\,,
\ee
so that $\Delta = 4ac-b^2 = -1 < 0$. The integral in Eq. (\ref{1.14}) is a 
standard well known integral whose answer depends on the sign of $a$ and 
$\Delta$. In our case $a > 0$ and $\Delta < 0$. Hence the integral in 
Eq. (\ref{1.14}) is given by \cite{gr}
\be\label{1.16}
\frac{1}{2\sqrt{(1-2m)}} \cosh^{-1}\left[\frac{4(1-2m)+(4m-1)u}{u}\right] = \sqrt{2}x \,. 
\ee
We thus have the solution
\be\label{1.17}
\cosh(t) = \frac{4(1-2m)+(4m-1)u}{u}\,,~~t = 2\sqrt{1-2m}x\,.
\ee
On inverting and noting that $u = 1+y = 1+\cn(\phi,m)$ we then have
the kink solution
\be\label{1.18}
\cn(\phi_{K},m) = \frac{(3-4m)\sech(t)-1}{1+(1-4m)\sech(t)}\,.
\ee
Thus as $x$ goes from $-\infty$ to +$\infty, \phi$ goes from $-2K(m)$ to 
+$2K(m)$ while at $x = 0, \phi$ goes to zero with an exponential kink tail.
The corresponding antikink solution goes from $\phi = 2K(m)$ to $\phi = 
-2K(m)$ as $x$ goes from $-\infty$ to $\infty$. Solution (25) is depicted in 
Figure 2 for m=1/4. 

\begin{figure}[h] 
\includegraphics[width=5.0 in]{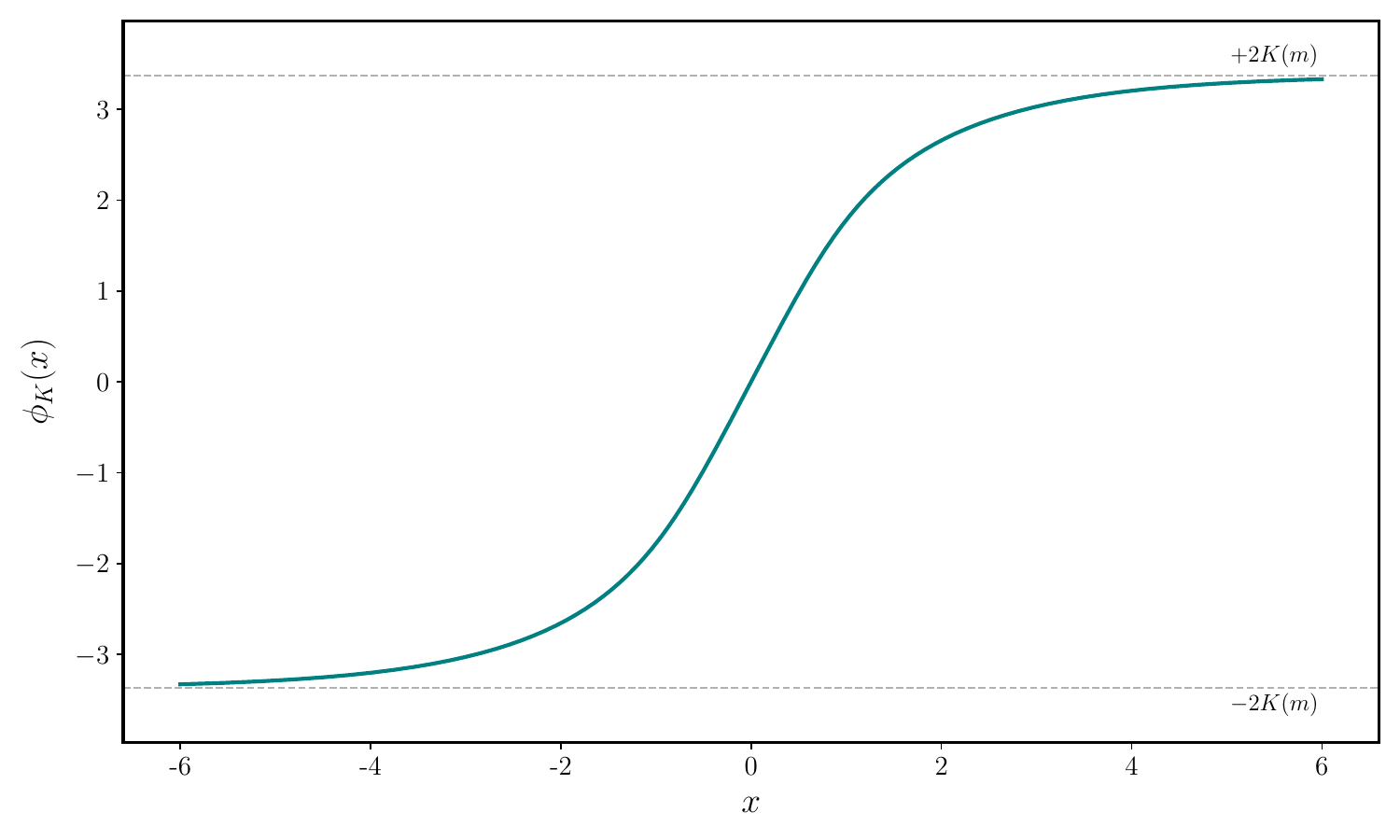}
\caption{Kink solution given by Eq. (25) for $m=1/4$. }
\end{figure}  

{\bf Kink Stability}

One can check the stability of the kink solution by considering the stability
equation (with linear perturbation $\eta$) 
\be\label{1.18a}
-\frac{d^2\eta}{dx^2} + V(x) \eta = \omega^2 \eta\,,
\ee
where
\be\label{1.18b}
V(x) = [V''(\phi)]_{\phi = \phi_{K}}\,.
\ee
On using Eqs. (\ref{1.7}), (\ref{1.17}) and (\ref{1.18}) the potential $V(x)$ 
turns out to be
\be\label{1.18c}
V(x) = \frac{(1-2m)[1-(3-4m)\sech(t)][1+(1-4m)\sech(t)][1+8(5-4m)m^2\sech(t)
(1-\sech(t))]}{1-2(1-8m+8m^2)\sech(t)+\sech^2(t)}\,,
\ee
where $t$ is as given by Eq. (\ref{1.17}). The corresponding zero mode 
eigenstates are 
\be\label{1.18e}
\omega_0 = 0\,,~~\eta_0 = \frac{d\phi_{K}(x)}{dx} \propto \frac{\sqrt{\sech(t)
[1+\sech(t)]}}{1-2(1-8m+8m^2)\sech(t)+\sech^2(t)}\,.
\ee
Potential given by Eq. (28) is depicted in Figure 3. As expected, $\eta_0$ is 
indeed nodeless and vanishes as $x \rightarrow \pm \infty$, thus indicating stability. 

\begin{figure}[h] 
\includegraphics[width=5.0 in]{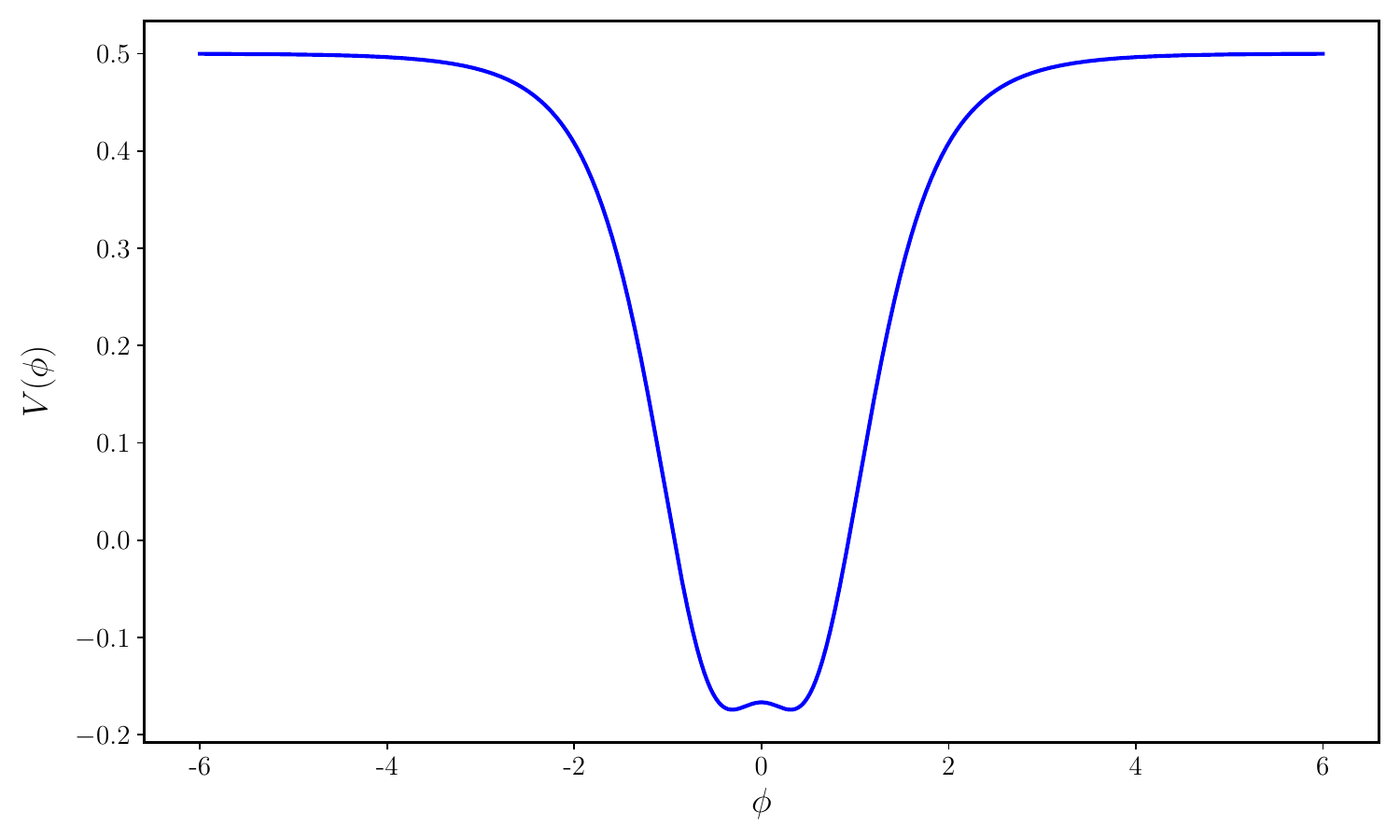}
\caption{Potential given by Eq. (28) for $m=1/4$. }
\end{figure}  

\subsection{Kink Solution at $m = 1/2$}

In this case we have 
\be\label{1.19}
\int \frac{du}{u\sqrt{2u-u^2}}  = x\,,
\ee
which is a well known integral giving \cite{gr}
\be\label{1.20}
u = 1+\cn(\phi,m) = \frac{2}{1+x^2}\,,
\ee
and hence 
\be\label{1.21}
\cn(\phi_{K},m) = \frac{(1-x^2)}{(1+x^2)}\,.
\ee
Thus we have a kink solution with a power law tail. As $x$ goes from $-\infty$ to
+$\infty, \phi$ goes from $-2K(m)$ to +$2K(m)$ while at $x = 0$, it goes to $0$.
The corresponding antikink solution goes from $\phi = 2K(m)$ to $\phi = 
-2K(m)$ as $x$ goes from $-\infty$ to $\infty$. The kink solution (32) for $m=1/2$ 
is depicted in Figure. 4. 

\begin{figure}[h] 
\includegraphics[width=5.0 in]{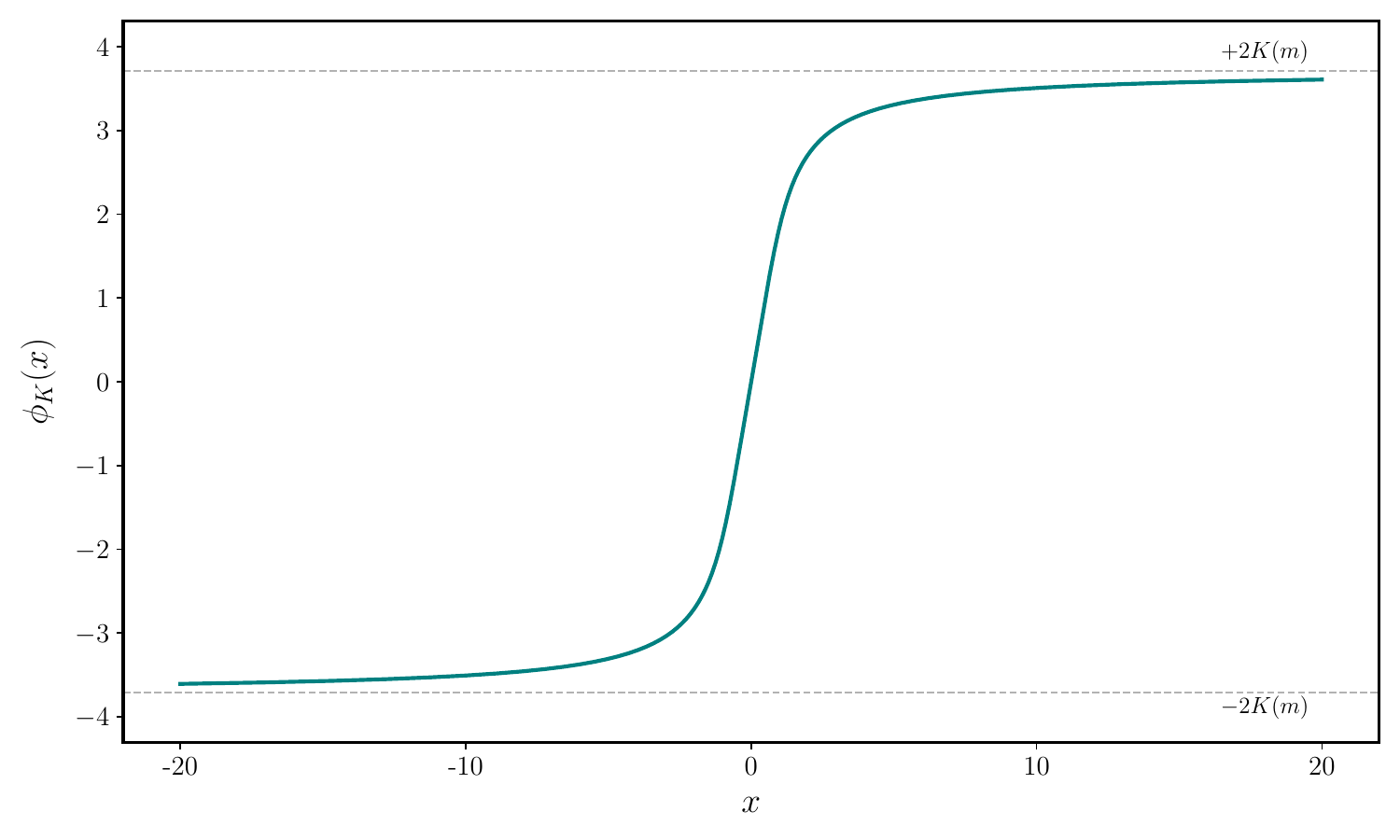}
\caption{Kink solution given by Eq. (32) for $m=1/2$. }
\end{figure}  

It is worth pointing out that so far there are very few analytically solvable
examples of kink solution with a power law tail \cite{ks4}. Thus the elliptic SG 
model with $m = 1/2$ is an interesting analytically exactly solvable example of 
having a kink solution with a power law tail. 

On using Eqs. (\ref{1.7}), (\ref{1.17}) and (\ref{1.21}) the potential $V(x)$ 
in the stability Eq. (\ref{1.18a}) turns out to be
\be\label{1.21c}
V(x) = \frac{6x^2(1-x^2)}{(1+x^4)^2}\,.
\ee
This potential is depicted in Figure 5. The corresponding zero mode eigenstates are 
\be\label{1.21d}
\omega_0 = 0\,,~~\eta_0 = \frac{d\phi_{K}(x)}{dx} \propto \frac{1}
{\sqrt{1+x^4}}\,.
\ee
As expected, $\eta_0$ is indeed nodeless and vanishes as $x \rightarrow \pm 
\infty$, thus indicating stability. 

\begin{figure}[h] 
\includegraphics[width=5.0 in]{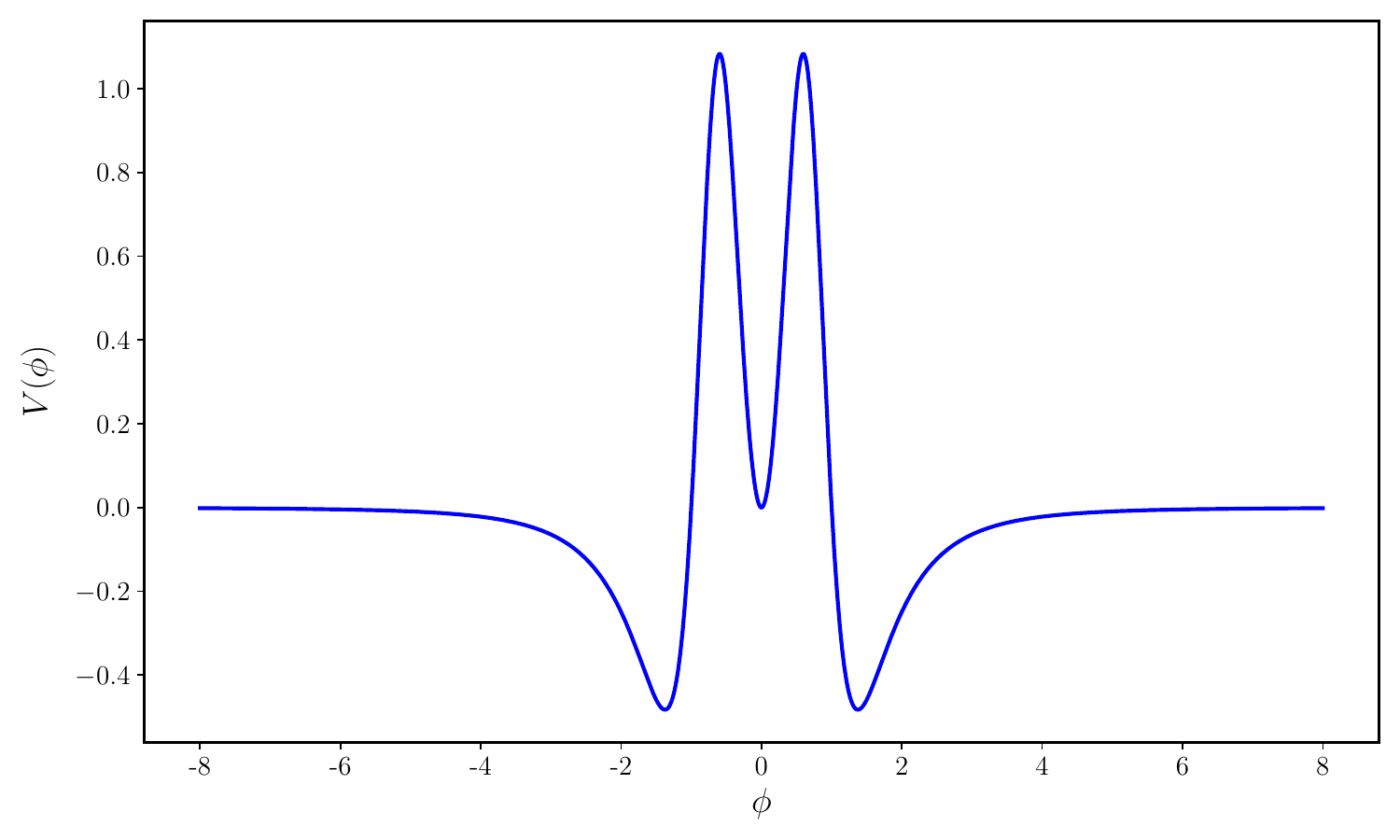}
\caption{Potential given by Eq. (33) for $m=1/2$. }
\end{figure}  

\subsection{Kink Solution For $1/2 < m  < 1$}

Let us now try to obtain the kink solution for $1/2 < m < 1$. In order to 
get the kink solution, we start from the static self-dual equation
\be\label{1.22}
\phi_{x} = \sqrt{2V(\phi)} 
= \pm \sqrt{2\left[\frac{\cn(\phi,m)}{\dn^2(\phi,m)} +\frac{1}{2\sqrt{m(1-m)}}\right]}\,,
\ee
where we have added a constant so that the potential is nonnegative. 
On using $\dn^2(x,m) = 1-m + m\cn^2(x,m)$, the integral is given by
\be\label{1.23}
\int \frac{\dn(\phi,m) d\phi}{\sqrt{[1-m +m\cn^2(\phi,m)+2\sqrt{m(1-m)}
\cn(\phi,m)]}} = \frac{x}{[m(1-m)]^{1/4}}\,.
\ee
On making the substitution $\cn(\phi,m) = y$, the integral becomes
\be\label{1.24}
\int \frac{dy}{(y+\sqrt{\frac{(1-m)}{m}}) \sqrt{(1-y^2)}} 
= x\left[\frac{m}{1-m}\right]^{1/4}\,.   
\ee
On substituting $y+\sqrt{\frac{(1-m)}{m}} = u$ we have the integral of the form
\be\label{1.25}
\int \frac{du}{u\sqrt{a+bu+cu^2}} = x\left[\frac{m}{1-m}\right]^{1/4}\,,  
\ee
where in this case
\be\label{1.26}
a = (2m-1)/m > 0\,,~~b = 2\sqrt{\frac{(1-m)}{m}}\,,~~c = -1\,,
\ee
so that $\Delta = 4ac-b^2 = -4 < 0$. The integral in Eq. (\ref{1.25}) is a standard 
well known integral \cite{gr} whose answer depends on the sign of $a$ and 
$\Delta$. In our case $a > 0$ and $\Delta < 0$. Hence, the integral in 
Eq. (\ref{1.25}) is given by
\be\label{1.27}
\frac{\sqrt{m}}{\sqrt{(2m-1)}} \cosh^{-1}\left[\frac{(2m-1)+\sqrt{m(1-m)}u}{m u}\right] 
= x\left[\frac{m}{1-m}\right]^{1/4}\,. 
\ee
We thus have the solution
\be\label{1.28}
\cosh(t) = \frac{(2m-1)+\sqrt{m(1-m)}u}{mu}\,,
~~t = \left[\frac{2m-1}{m(1-m)}\right]^{1/4} x\,.
\ee
On inverting and noting that $u = 1+y = 1+\cn(\phi,m)$ we then have
\be\label{1.29}
\cn(\phi_{K},m) = \frac{\sqrt{m}\sech(t)-\sqrt{1-m}}
{\sqrt{m}-\sqrt{1-m}\sech(t)}\,.
\ee
Thus as $x$ goes from $-\infty$ to +$\infty, \phi$ goes from $-\phi_c$ to 
+$\phi_c$ while at $x = 0, \phi$ goes to zero. Here, as defined above
$\cn(\pm \phi_c,m) = -\sqrt{\frac{(1-m)}{m}}$. Thus we have a kink solution
with an exponential tail. The kink solution (42) for $m=3/4$ is depicted in Figure 6. 
The corresponding antikink solution goes from $\phi = \phi_c$ to $\phi = 
-\phi_c$ as $x$ goes from $-\infty$ to $\infty$.

\begin{figure}[h] 
\includegraphics[width=5.0 in]{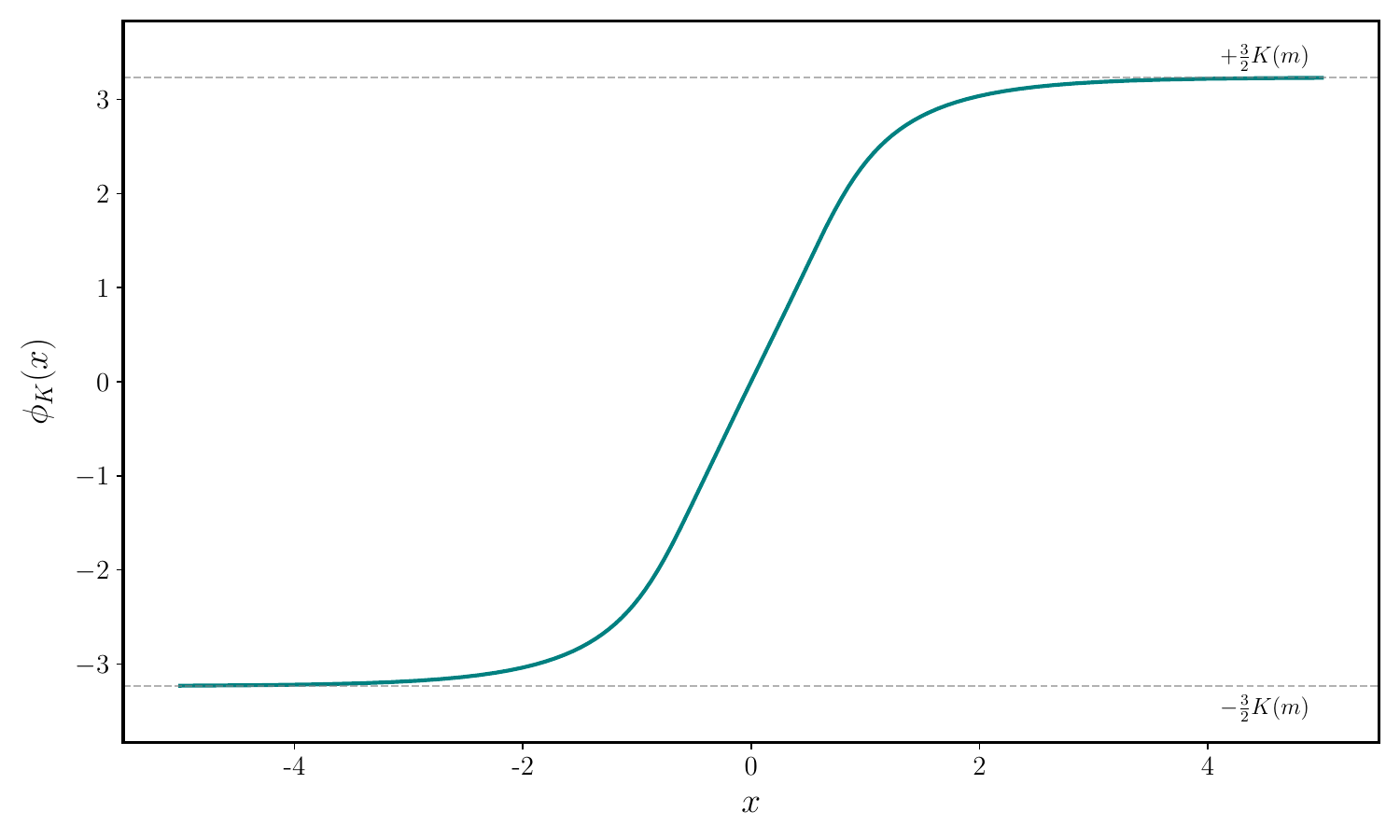}
\caption{Kink solution given by Eq. (42) for $m=3/4$. }
\end{figure}  

On using Eqs. (\ref{1.7}), (\ref{1.17}) and (\ref{1.29}) the potential $V(x)$ 
in the stability Eq. (\ref{1.18a}) turns out to be
\be\label{1.29a}
V(x) = \frac{N(x)}{D(x)}\,,
\ee
where
\bea\label{1.29b}
&&N(x) = (2m-1)[\sqrt{m(1-m)}(1+\sech^2(t))-\sech(t)] [4m(1-m) \nonumber \\
&&+2\sqrt{m(1-m)}\sech(t)+(4m(1-m)-1)\sech^2(t)]\,,
\eea
\be\label{1.29c}
D(x) = [2m(1-m)-2\sqrt{m(1-m)}\sech(t)+(1-2m+2m^2)\sech^2(t)]^2\,,
\ee
and $t$ is as given by Eq. (\ref{1.28}). The potential given by  Eqs. (43) 
to (45) is depicted in Fig. 7. 

\begin{figure}[h] 
\includegraphics[width=5.0 in]{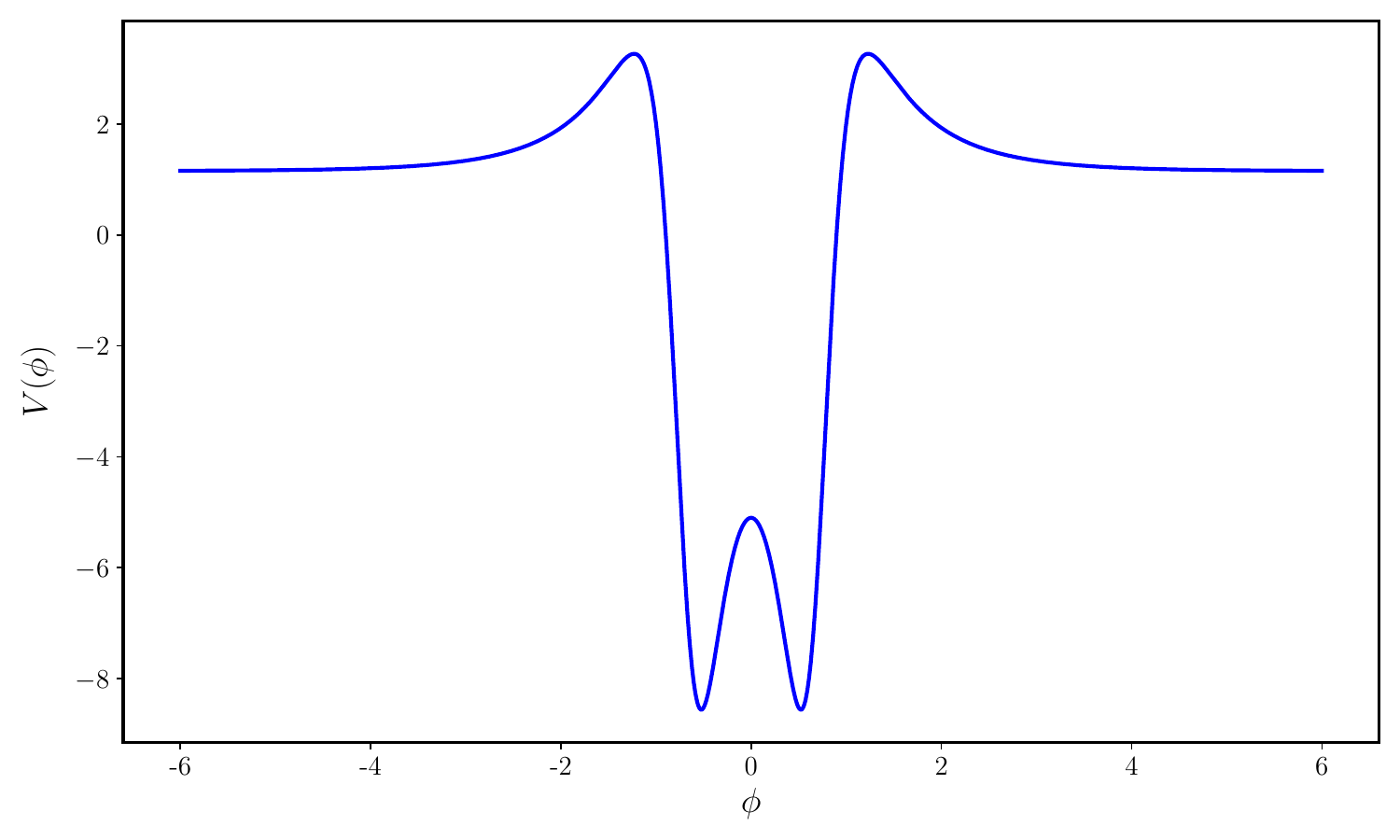}
\caption{Potential given by Eqs. (43) to (45) for $m=3/4$. }
\end{figure}  

The corresponding zero mode eigenstates are 
\be\label{1.29d}
\omega_0 = 0\,,~~\eta_0 = \frac{d\phi_{K}(x)}{dx} \propto \frac{\sech(t)}
{\sqrt{2m(1-m)-2\sqrt{m(1-m)}\sech(t)+(1-2m+2m^2)\sech^2(t)}}\,.
\ee
As expected, $\eta_0$ is indeed nodeless and vanishes as $x \rightarrow \pm 
\infty$.  Summarizing, we have an exponential kink tail for all values of 
$0 \le m < 1$ except for $m = 1/2$ when one has a power law kink tail.  

\section{Summary And Some Open Problems}

In this paper we have presented a continuous one-parameter family of elliptic
sine-Gordon equation characterized by the parameter $0 \le m \le 1$ and 
obtained kink and antikink solutions for all values of $0 < m < 1$. This is an 
interesting example which in both the extreme limits of $m = 0$ and $m = 1$ 
goes over to integrable SG and SHG equations, respectively. For this model we 
showed that the kink solutions have 
exponential tails except when $m = 1/2$ in which case we have a power law tail.  
It is worth pointing out that there are very few well-known examples \cite{ks4} of 
analytically solvable kink solutions with a power law tail so this is a valuable 
addition to that list. This paper raises several questions some of which are: 

\begin{enumerate}

\item Since the elliptic SG model goes over to the integrable models in both
the extreme limits of $m = 0$ and $m = 1$ where it goes over to the SG and SHG 
respectively, hence it is worth enquiring if this is an integrable model for 
all values of $0 \le m \le 1$. Most likely the answer would be no, but it is 
important to know the definite answer.

\item In case it is not an integrable model for $0 < m < 1$, then the next 
question is how 
close $m$ should be to $0$ as well as to $m = 1$ to see the effect of 
integrability? More precisely, how well is the constancy of the formerly 
conserved quantities preserved in the nonintegrable limit of $m \ne 0,1$?
Recently such a question was posed and discussed in some detail in the 
context of the discrete nonintegrable Salerno model vis a vis the integrable 
Ablowitz-Ladik model \cite{mit}.

\item Can one analytically obtain a few other solutions of the elliptic SG 
model such as the pulse solutions and even the complex PT-invariant solutions?  

\item While the kink-kink and the kink-antikink forces are easily calculated in
this model (except when $m = 1/2$) using Manton's formula \cite{man}, the 
calculation of the kink-kink and the kink-antikink force at $m = 1/2$ is an 
interesting open
problem \cite{man1,pan} since so far no definitive answer has been found in 
case the kink solutions have a power law tail \cite{ks4}.

\item Finally, can one think of other models (with Jacobi elliptic functions  
or not) which are integrable in the two extreme limits and then study various 
features of such a model?

\end{enumerate}

We hope to address some of these issues in the near future. \\ 

\noindent {\bf Acknowledgment} We are indebted to Robin Msiska for the help with the 
figures. One of us AK is grateful to the Indian National Science Academy 
(INSA) for the award of INSA Honorary Scientist position at Savitribai
Phule Pune University.  The work at Los Alamos National Laboratory  was supported 
by the US Department of Energy.


\begin{thebibliography}{99}

\bibitem{raj} See for example, R. Rajaraman, Solitons and Instantons: An 
Introduction to Solitons and Instantons in Quantum Field Theory (Elsevier, 
1982). 

\bibitem{cou} See for example, The Sine Gordon Model and its Applications, 
From Pendula and Josephson Junctions to Gravity and High Energy Physics, 
Edited by J. Cuevas-Maravar, P. G. Kevrekidis and F. Williams (Springer, 2014).

\bibitem{shg} H. Zhang, Chaos, Solitons \& Fractals {\bf 28} (2006) 489. 

\bibitem{as} See For example, M. Abramowitz and I.A. Stegun, Handbook of 
Mathematical Functions (Dover, 1964).

\bibitem{gr} I. S. Gradshteyn and I. M. Ryzhik, Tables of Integrals, Series and 
Products, (Academic Press, New York, 2007). 

\bibitem{ks4} A. Khare and A. Saxena, Frontiers in Physics {\bf 10} (2022)
992915.

\bibitem{mit} T. Mithun, A. Maluckov, A. Mancic, A. Khare and P.G. Kevrekidis,
Phys. Rev. E {\bf 107} (2023) 004202. 

\bibitem{man} N.S. Manton, Nucl. Phys. B {\bf 150} (1979) 397. 

\bibitem{man1} N.S. Manton, J. Phys. A {\bf 52} (2019) 065401. 

\bibitem{pan} I. C. Christov, R. J. Decker, A. Demirkaya, V. K. Gani, P. G.  
Kevrekidis, A. Khare and A. Saxena, Phys. Rev. Lett. {\bf 122} (2019) 171601.

\end{thebibliography}
\end{document}